\documentclass[aps,prd,showpacs,twocolumn]{revtex4}%
\usepackage{amssymb}
\usepackage{amsfonts}
\usepackage{amsmath}
\usepackage{graphicx}
\usepackage[usenames,dvipsnames]{color}%
\providecommand{\U}[1]{\protect\rule{.1in}{.1in}}
\begin{document}
\title{Generation of geometrical phases\textbf{\ }and persistent spin
currents\textbf{\ }in $1$-dimensional rings by Lorentz-violating terms}
\author{R. Casana, M. M. Ferreira Jr, V. E. Mouchrek-Santos, Edilberto O. Silva}
\affiliation{Universidade Federal do Maranh\~{a}o (UFMA), Departamento de F\'{\i}sica,
Campus Universit\'{a}rio do Bacanga, S\~{a}o Lu\'{\i}s - MA, 65080-805 - Brazil}

\begin{abstract}
We have demonstrated that Lorentz-violating terms stemming from the fermion
sector of the SME are able to generate geometrical phases on the wave function
of electrons confined in 1-dimensional rings, as well as persistent spin
currents, in the\ total absence of electromagnetic fields. We have explicitly
evaluated the eigenenergies and eigenspinors of the electrons modified by the
Lorentz-violating terms, using them to calculate the dynamic and the
Aharonov-Anandan phases in the sequel. The total phase presents a pattern very
similar to the Aharonov-Casher phase accumulated by electrons in rings under
the action of the Rashba interaction. Finally, the persistent spin current
were carried out and used to impose upper bounds on the Lorentz-violating parameters.

\end{abstract}

\maketitle

\section{Introduction}

The standard model extension (SME) \cite{Colladay} was proposed as an
extension of the minimal standard model including terms of Lorentz symmetry
violation in all interaction sectors. Its gauge sector has been much
investigated in several respects \cite{Jackiw,Mewes,Soldati,Adam,Baeta},
including photon-fermion interactions \cite{Vertex}, nonminimal couplings with
higher order derivatives \cite{HD}, and higher dimension operators
\cite{Reyes}. The initial investigations on the fermion sector of the SME were
associated with the breaking of the CPT symmetry \cite{Fermion1}, its
consistency, stability, hermiticity, quantization respects \cite{Fermion2},
its nonrelativistic regime, Foldy-Wouthuysen transformation \cite{Fermion3},
and some modified Dirac equations \cite{DiracE}. The fermion sector of the
SME\ is described by the following Lagrangian,
\begin{equation}
\mathcal{L}_{Total}=\frac{1}{2}i\overline{\psi}\Gamma^{\nu}%
\overleftrightarrow{\partial}_{\nu}\psi-\overline{\psi}M\psi,
\end{equation}
where the field $\psi$\ is a Dirac spinor, and%
\begin{align}
\Gamma^{\nu} &  =\gamma^{\nu}+c^{\mu\nu}\gamma_{\mu}+d^{\mu\nu}\gamma
_{5}\gamma_{\mu}+\frac{1}{2}g^{\lambda\mu\nu}\sigma_{\lambda\mu},\label{GF1}\\
M &  =m+a_{\mu}\gamma^{\mu}+b_{\mu}\gamma_{5}\gamma^{\mu}-\frac{1}{2}H_{\mu
\nu}\sigma^{\mu\nu},\label{MF1}%
\end{align}
with the terms $c^{\mu\nu},d^{\mu\nu},g^{\lambda\mu\nu},$ $a_{\mu},b_{\mu}%
$\ and $H_{\mu\nu}$\ standing for Lorentz-violating (LV) tensors, which have
the following mass dimension: $\left[  a_{\mu}\right]  =\left[  b_{\mu
}\right]  =\left[  H_{\mu\nu}\right]  =1,$ $\left[  c^{\mu\nu}\right]
=\left[  d^{\mu\nu}\right]  =\left[  g^{\lambda\mu\nu}\right]  =0.$ While
$a_{\mu}$\ and $b_{\mu}$\ break the CPT symmetry, the terms $c^{\mu\nu}%
,d^{\mu\nu}$\ and $H_{\mu\nu}$\ are CPT-even. The modified Dirac equation is
\begin{equation}
\left[  i\Gamma^{\nu}\partial_{\nu}-M\right]  \psi=0,
\end{equation}
whose nonrelativistic regime was analyzed in Ref. \cite{Fermion3}. The
physical effects induced by these parameters are used to state upper bounds on
their magnitude \cite{KostelecR}, \cite{AltschulR}.

Condensed matter systems constitute a proper environment in which
Lorentz-violating theories may find application, once are usually endowed with
rotation invariance breakdown or privileged directions. Two dimensional
electron systems have demonstrated to be a rich field where spin-orbit
interaction plays a very relevant role in connection with the spintronics,
where geometrical phases and persistent currents are remarkable observables.
Geometrical phases in quantum mechanics have been much investigated since
Berry's seminal demonstration about the phase accumulated in a cyclic
adiabatic evolution \cite{Berry}, and the Aharonov-Anandan's discovery about
the geometrical phase developed in a cyclic nonadiabatic evolution \cite{AA}.
Observable effects of geometrical phases in condensed matter systems have been
reported since the 80's, as the conductance oscillations due to Aharonov-Bohm
effect in mesoscopic systems \cite{Aronov1}. In the early 90's, it was shown
that the motion of electrons in mesoscopic rings in the presence of magnetic
field implies the generation of Berry phase \cite{Loss}. In 1992, Mathur \&
Stone \cite{Mathur} showed that conductance oscillations in semiconductors are
a signature of the Aharonov-Casher effect \cite{AC}. Balatsky \& Altshuler
\cite{Balatsky} investigated the motion of electrons in one-dimensional rings
under the action of the electric field responsible by the Aharonov-Casher
effect. Aronov \& Lyanda-Geller \cite{Aronov2} investigated the motion of
electron in conducting rings, showing that the spin-orbit interaction gives
origin to a Berry phase in an adiabatic cyclic evolution. Qian \& Su
\cite{Qian} argued that electrons in a nonadiabatic cyclic evolution in
mesoscopic rings acquire a Aharonov-Casher phase composed of a dynamic and an
Aharonov-Anandan phase, which in the adiabatic limit recovers the Berry phase
obtained by Aronov \& Lyanda-Geller. Implications involving the geometrical
phases associated with persistent currents \cite{Oh} and transport properties
in mesoscopic rings \cite{Choi} were addressed in several works
\cite{Loss,Balatsky,Molnar,Meijer,Zhou,Nitta,Berche,Gigl,Splett,Shen,Frus,Citro,Kovalev}%
.

The Rashba interaction on electrons confined in a 1-dimensional ring can be
addressed as stated in Ref. \cite{Choi}. The starting Hamiltonian is
$H=\frac{1}{2m}\left[  \mathbf{p}-\mu\left(  \boldsymbol{\sigma}%
\times\boldsymbol{E}\right)  \right]  ^{2},$ where the electric field is
$\mathbf{E}=E\left(  \cos\chi\hat{r}-\sin\chi\hat{z}\right)  ,$ the term
$\mathbf{p}\cdot\left(  \boldsymbol{\sigma}\times\boldsymbol{E}\right)
$\ represents the Rashba interaction, and $\boldsymbol{\sigma=(}\sigma
_{x},\sigma_{y},\sigma_{z})$ represents the Pauli matrices. The corresponding
eigenenergies are,
\begin{equation}
E=\frac{1}{2mr_{0}^{2}}\left[  n-\frac{\Phi_{AC}}{2\pi}\right]  ^{2}%
,\label{Echoi}%
\end{equation}
where $r_{0}$ is the radius of the ring, and $\Phi_{AC}=-\pi\left(
1-\lambda_{\pm}\right)  $ is the Aharonov-Casher phase developed by the
eigenspinor,
\begin{equation}
\Psi^{\left(  \pm\right)  }=e^{in\varphi}%
\begin{bmatrix}
\cos\left(  \beta_{\pm}/2\right) \\
\pm\sin\left(  \beta_{\pm}/2\right)  e^{i\varphi}%
\end{bmatrix}
,\label{Spchoi}%
\end{equation}
after a complete cycle. Here, $\lambda_{\pm}=\pm\sqrt{\omega_{1}^{2}+\left(
\omega_{3}+1\right)  ^{2}},$ $\beta_{-}=\pi-\beta_{+}$, $\tan\beta_{\pm
}=\omega_{1}/\left(  \omega_{3}+1\right)  ,$ $\omega_{1}=\left(  \mu
Er/\hslash c\right)  \sin\chi,$ $\omega_{3}=\left(  \mu Er/\hslash c\right)
\cos\chi.$ From the eigenspinors (\ref{Spchoi}), one evaluates a dynamic
phase, $\Phi_{dyn}^{\left(  \pm\right)  },$ and the Aharonov-Anandan
geometrical phase, $\Phi_{AA}^{\left(  \pm\right)  }$, whose sum yields the
total AC phase that appears in Eq. (\ref{Echoi}).

Investigations about the generation of topological phases by Lorentz-violating
terms have begun in the context of a CPT-odd nonminimal coupling between
photons and fermions \cite{Nonminimal}. New studies involving Aharonov-Bohm,
Aharonov-Casher and topological phases in scenarios endowed with Lorentz
symmetry breaking were developed in Refs. \cite{Nonminimal2,Knut}. The
interesting similarity between Lorentz-violating terms of the SME fermion
sector, regarded in the nonrelativistic limit, and some spin-orbit
interactions of the condensed matter systems, as the Rashba term,
$H_{Rashba}=\alpha_{R}\left(  \sigma_{y}p_{x}-\sigma_{x}p_{y}\right)  ,$ and
the Dresselhauss interaction, $H_{D}=\beta\left(  \sigma_{x}p_{x}-\sigma
_{y}p_{y}\right)  ,$ were discussed in recent works
\cite{Ajaib,NM3,Bakke,Ajaib2}. It was verified that the nonrelativistic limit
of the mass term, $H_{\mu\nu}\sigma^{\mu\nu},$ leads to the Rashba spin-orbit
interaction, that appears as a consequence of an inversion asymmetry potential
in a semiconductor interface, in the presence of an electric field
\cite{Berche}. In this case, the tensor component $H_{0i}$ corresponds to the
Rashba\ coupling constant ($\alpha_{R}$): $\alpha_{R}=H_{0i}/m$, so that
$H_{0i}$ plays the role of the electric field, $E$\textbf{. }The Rashba term
has appeared also in the context of the Dirac equation modified by a CPT-even
nonminimal coupling \cite{NM3} and a CPT-odd nonminimal coupling \cite{Bakke}.
\ Recently, it was discussed that the term $d^{\mu\nu}$ can recover the
Dresselhauss interaction \cite{Ajaib2}.

The purpose of this work is to show that the tensor background, $d^{\mu\nu} $,
provides nonrelativistic contributions to the Hamiltonian of electrons
confined in a 1-dimensional ring, which alter the corresponding eigenenergies
and eigenspinors in a compatible way with the generation of geometrical phases
analogue to the ones produced by the Rashba interaction in condensed matter
systems. It occurs in the entire absence of electric or magnetic fields. These
phases are also associated with induced persistent spin currents, which
constitute a feasible route to constrain the Lorentz-violating parameters in
mesoscopic systems at a level much better than previously supposed
\cite{Ajaib2}.

\section{Induction of geometrical phases and persistent spin currents by
Lorentz-violating terms}

We now investigate the effects played by some terms stemming from the fermion
sector of the SME on the wave function of electrons confined in 1-dimensional
rings, pointing out that we are using natural units, $\hslash=1,c=1.$
Particularly, we are interested in the Dirac equation, in the absence of
electromagnetic field,
\begin{equation}
\left(  i\Gamma^{\nu}\partial_{\nu}-m\right)  \Psi=0,\label{Dirac1}%
\end{equation}
where
\begin{equation}
\Gamma^{\nu}=\gamma^{\nu}+d^{\mu\nu}\gamma_{5}\gamma_{\mu},
\end{equation}
with $d^{\mu\nu}$ being the CPT-even, symmetric and traceless tensor belonging
to fermion sector of the SME. The nonrelativistic Hamiltonian associated with
Eq. (\ref{Dirac1}), obtained under the condition $\left\vert \mathbf{p}%
\right\vert ^{2}<<m^{2},$ is \textbf{\ }%
\begin{align}
H  & =\frac{\mathbf{p}^{2}}{2m}+md_{j0}\sigma^{j}+d_{jk}p^{j}\sigma^{k}%
+d_{00}p^{j}\sigma^{j}\\
& -\frac{3}{2}d_{0j}\frac{p_{j}p^{l}\sigma^{l}}{m}+\frac{\mathbf{p}^{2}%
}{2m^{2}}d_{mj}p^{j}\sigma^{m}+d_{jk}\frac{p_{j}p_{k}}{2m^{2}}p^{l}\sigma
^{l}.\nonumber
\end{align}
The effective nonrelativistic Hamiltonian, composed of the more meaningful
terms, is
\begin{equation}
H=\frac{\mathbf{p}^{2}}{2m}-md^{j}\sigma^{j}+d_{jk}p^{j}\sigma^{k}+d_{00}%
p^{j}\sigma^{j},\label{Hdij2}%
\end{equation}
where $d^{j}=-d_{j0}.$ The term $d^{j}\sigma^{j}=\mathbf{d}\cdot
\boldsymbol{\sigma}$ acts providing a Zeeman-like effect\ or a magnetic moment
contribution in the case the vector $\mathbf{d}$ could play the role of a
magnetic field. This term will not be considered anymore, since we focus
attention on the terms $d_{jk}p^{j}\sigma^{k}$ and $d_{00}p^{j}\sigma^{j}$
that provide analogue structures to the Rashba and Dresselhauss interactions.
The tensor $d^{\mu\nu}$ is $CPT-$even, and its elements can be classified
under the action of the discrete operations:\textbf{\ }$P$ (parity)$,$ $C$
(charge conjugation)$,$ $T$ (time reversal). All them are $C-$odd, and
$PT-$odd. The elements $d^{00},d^{ij}\ $are $T-$even and $P-$odd, while the
coefficients $d^{0i}$ are $T-$odd and $P-$even. The T-even character of
$d^{00},d^{ij}$ will allow to obtain persistent spin current but no charge
current \cite{Oh,Zhou,Berche} in the next sections.

\subsection{Phases and spin currents generated by the coefficients $d^{ij}$}

We begin our investigation doing $d^{00}=d^{0i}=0,$ and keeping our attention
in the term $d_{jk}p^{j}\sigma^{k}$, so that the Hamiltonian (\ref{Hdij2})
becomes
\begin{equation}
H=\frac{\mathbf{p}^{2}}{2m}+d^{ij}\sigma^{i}p^{j},\label{Hdij3}%
\end{equation}
presenting a general structure analogue to the Rashba and Dresselhauss
interactions. The term $d^{ij}\sigma^{i}p^{j}$ can be explicitly written as
\begin{equation}
d^{ij}\sigma^{i}p^{j}=d^{11}\sigma_{x}p_{x}+d^{22}\sigma_{y}p_{y}%
+d^{12}\left(  \sigma_{x}p_{y}+\sigma_{y}p_{x}\right)  .
\end{equation}
Here, we have used $d^{12}=d^{21}$, and taken $p_{z}=0$, once the electron
moves on the plane. We have also set $d^{32}=d^{31}=d^{33}=0,$ so that
$d^{ij}$ becomes a $2\times2$ matrix. Choosing $d^{22}=-d^{11}$, the tensor
$d^{\mu\nu}$ becomes traceless, as required. Considering that the electron
moves on a ring of fixed radius, $r=r_{0},$ we can write the momentum in polar
coordinates, \textbf{\ }%
\begin{equation}
p_{x}=\frac{\sin\varphi}{r_{0}}\left[  i\frac{\partial}{\partial\varphi
}\right]  ,\text{ \ }p_{y}=-\frac{\cos\varphi}{r_{0}}\left[  i\frac{\partial
}{\partial\varphi}\right]  ,
\end{equation}
so that
\begin{align}
d^{ij}\sigma^{i}p^{j}  & =\alpha_{11}\left(  \sigma_{x}\sin\varphi+\sigma
_{y}\cos\varphi\right)  \left[  i\frac{\partial}{\partial\varphi}\right] \\
& +\alpha_{12}\left(  -\sigma_{x}\cos\varphi+\sigma_{y}\sin\varphi\right)
\left[  i\frac{\partial}{\partial\varphi}\right]  ,\nonumber
\end{align}
where $\ \alpha_{11}=\left(  1/r_{0}\right)  d^{11},$ \ $\alpha_{12}=\left(
1/r_{0}\right)  d^{12}.$

The Hamiltonian (\ref{Hdij3}), in polar coordinates, is
\begin{align}
H  & =\Omega\left[  i\frac{\partial}{\partial\varphi}\right]  ^{2}+\alpha
_{11}\left(  \sigma_{x}\sin\varphi+\sigma_{y}\cos\varphi\right)  \left[
i\frac{\partial}{\partial\varphi}\right] \nonumber\\
& +\alpha_{12}\left(  -\sigma_{x}\cos\varphi+\sigma_{y}\sin\varphi\right)
\left[  i\frac{\partial}{\partial\varphi}\right]  ,\label{Hdij4}%
\end{align}
with $\Omega=1/2mr_{0}^{2}$. We can observe that the term $d^{ij}\sigma
^{i}p^{j}$ yields a contribution equal to the Dresselhauss interaction,
\begin{equation}
H_{Dresselhaus}=\alpha_{D}\left(  \sigma_{x}\sin\varphi+\sigma_{y}\cos
\varphi\right)  \left[  i\frac{\partial}{\partial\varphi}\right]  ,
\end{equation}
and one analogue, but not equal, to the Rashba interaction,\textbf{\ }%
\begin{equation}
H_{Rashba}=\alpha_{R}\left(  \sigma_{y}\sin\varphi+\sigma_{x}\cos
\varphi\right)  \left[  i\frac{\partial}{\partial\varphi}\right]  .
\end{equation}
When\ the Hamiltonian (\ref{Hdij3}) is written in polar coordinates, as Eq.
(\ref{Hdij4}), it lets to be hermitian, as pointed out in Refs.
\cite{Meijer,Berche}. In order to find its hermitian form, we follow the
procedure of Ref. \cite{Berche}, achieving
\begin{align}
H  & =\Omega\left[  i\frac{\partial}{\partial\varphi}\right]  ^{2}+\left[
\alpha_{11}\left(  \tilde{\sigma}_{\rho}\right)  -\alpha_{12}\left(
\tilde{\sigma}_{\varphi}\right)  \right]  \left[  i\frac{\partial}%
{\partial\varphi}\right] \nonumber\\
& +\frac{i}{2}\left[  \alpha_{11}\left(  \tilde{\sigma}_{\varphi}\right)
+\alpha_{12}\left(  \tilde{\sigma}_{\rho}\right)  \right]  ,\label{H.0.19}%
\end{align}
where%
\begin{align}
\tilde{\sigma}_{\rho} &  =\sigma_{x}\sin\varphi+\sigma_{y}\cos\varphi
,\label{H.0.14.a}\\
\tilde{\sigma}_{\varphi} &  =\sigma_{x}\cos\varphi-\sigma_{y}\sin
\varphi,\label{H.0.14.b}%
\end{align}
with $\tilde{\sigma}_{\varphi}=\partial_{\varphi}\tilde{\sigma}_{\rho},$
$\tilde{\sigma}_{\rho}=-\partial_{\varphi}\tilde{\sigma}_{\varphi},$ and\
\begin{equation}
\ \tilde{\sigma}_{\rho}=%
\begin{bmatrix}
0 & -ie^{-i\varphi}\\
ie^{i\varphi} & 0
\end{bmatrix}
,\text{ \ }\tilde{\sigma}_{\varphi}=%
\begin{bmatrix}
0 & e^{i\varphi}\\
e^{-i\varphi} & 0
\end{bmatrix}
.
\end{equation}

Except for a constant term, $\left(  \alpha_{11}^{2}+\alpha_{12}^{2}\right)
/\left(  2\hslash\Omega\right)  ^{2}$, the Hamiltonian (\ref{H.0.19}) can be
compactly read as
\begin{equation}
H=\Omega\left[  i\frac{\partial}{\partial\varphi}+\frac{1}{2\Omega}\left[
\alpha_{11}\left(  \tilde{\sigma}_{\rho}\right)  -\alpha_{12}\left(
\tilde{\sigma}_{\varphi}\right)  \right]  \right]  ^{2},\label{H.0.20}%
\end{equation}
the form to be considered henceforth.

Now, we should evaluate the eigenenergies of electrons governed by the
nonrelativistic Hamiltonian (\ref{H.0.20}). We solve this problem for
$H^{\prime}$, given as
\begin{equation}
H^{\prime}=\left[  i\frac{\partial}{\partial\varphi}+\left[  \beta_{11}\left(
\tilde{\sigma}_{\rho}\right)  -\beta_{12}\left(  \tilde{\sigma}_{\varphi
}\right)  \right]  \right]  ,\label{H.2.a}%
\end{equation}
that is, $H^{\prime}\Psi=\Lambda\Psi,$ so that the eigenenergies are
$E=\Omega\Lambda^{2}$. Here, $\beta_{12}=\alpha_{12}/2\Omega,\ \beta
_{11}=\alpha_{11}/2\Omega.$ This operator has the matrix form,
\begin{equation}
H^{\prime}=\left[
\begin{array}
[c]{cc}%
i\frac{\partial}{\partial\varphi} & -\beta e^{i\left(  \varphi+\delta\right)
}\\
-\beta e^{-i\left(  \varphi+\delta\right)  } & i\frac{\partial}{\partial
\varphi}%
\end{array}
\right]  ,\label{H.4}%
\end{equation}
where%
\begin{equation}
\beta=\sqrt{\beta_{12}^{2}+\beta_{11}^{2}}\text{, \ }\tan\delta=\frac
{\beta_{11}}{\beta_{12}}\text{,}\label{H.4.1}%
\end{equation}
or%
\begin{equation}
\mathbf{\ \ }\beta=mr_{0}\sqrt{d_{12}^{2}+d_{11}^{2}}.\label{betaLV2}%
\end{equation}

The eigenspinors have the general form,
\begin{equation}
\Psi_{n,\lambda}\left(  \varphi\right)  =e^{i\lambda n\varphi}\tilde{\Psi
}\left(  \varphi\right)  =e^{i\lambda n\varphi}\left[
\begin{array}
[c]{c}%
a\\
be^{-i\varphi}%
\end{array}
\right]  ,\label{H.4.2}%
\end{equation}
with $n\in\mathbb{Z}$, $\lambda=\pm1.$ Solving the equation, $H^{\prime}%
\Psi\left(  \varphi\right)  =\Lambda\Psi\left(  \varphi\right)  ,$ we find
eigenenergies
\begin{equation}
E_{n}=\Omega\left\{  \lambda n-\frac{1}{2}\left[  1+\left(  -1\right)  ^{\mu
}\sqrt{1+4\beta^{2}}\right]  \right\}  ^{2},
\end{equation}
associated with spin down $\left(  \mu=1\right)  $ or spin up $\left(
\mu=2\right)  $ states. This energy expression can be read as%
\begin{equation}
E_{n}=\Omega\left[  \lambda n-\left(  -1\right)  ^{\mu}\frac{\Phi
_{Total}^{\left(  \mu\right)  }}{2\pi}\right]  ^{2},\label{H.14}%
\end{equation}
where
\begin{equation}
\Phi_{Total}^{\left(  \mu\right)  }=\left(  -1\right)  ^{\mu}\pi\left[
1+\left(  -1\right)  ^{\mu}\sqrt{1+4\beta^{2}}\right]  ,\label{Total1}%
\end{equation}
is the total phase induced on the electron eigenspinors,
\begin{equation}
\Psi_{n,\lambda}^{\left(  \mu\right)  }=e^{i\lambda n\varphi}\tilde{\Psi
}^{\left(  \mu\right)  }\left(  \varphi\right)  ,\label{Spinor1a}%
\end{equation}
by the Lorentz-violating term, analogue to the Aharonov-Casher phase induced
by the Rashba term \cite{Qian,Oh,Choi,Molnar,Frus,Berche}. This total phase is
composed of a dynamic part and a geometrical contribution, as it will be seen
as follows.

The normalized eigenspinors, $\tilde{\Psi}^{\left(  \mu\right)  }\left(
\varphi\right)  ,$ corresponding to the eigenenergies (\ref{H.14}),
\begin{align}
\tilde{\Psi}^{\left(  1\right)  }\left(  \varphi\right)   &  =\left[
\begin{array}
[c]{c}%
\cos\left(  \theta/2\right) \\
\sin\left(  \theta/2\right)  e^{-i\left(  \delta+\varphi\right)  }%
\end{array}
\right]  ,\label{H.19.a}\\
\tilde{\Psi}^{\left(  2\right)  }\left(  \varphi\right)   &  =\left[
\begin{array}
[c]{c}%
\sin\left(  \theta/2\right) \\
-\cos\left(  \theta/2\right)  e^{-i\left(  \delta+\varphi\right)  }%
\end{array}
\right]  ,\label{H.19.b}%
\end{align}
represent the spin down and spin up, respectively. Here, we also have
\begin{equation}
\tan\left(  \theta/2\right)  =-\frac{1}{\beta}\left[  \frac{1}{2}-\frac{1}%
{2}\sqrt{1+4\beta^{2}}\right]  ,\label{tanLV}%
\end{equation}%
\begin{equation}
\sin^{2}\left(  \theta/2\right)  =\left(  1-\sqrt{4\beta^{2}+1}\right)
^{2}\Xi,\label{H.21.1a}%
\end{equation}%
\begin{align}
\sin\theta & =-4\beta\left(  1-\sqrt{4\beta^{2}+1}\right)  \Xi,\label{H.23}\\
\cos\theta & =1/\sqrt{1+4\beta^{2}},\label{H.23b}%
\end{align}
where $\Xi=\left[  \left(  \sqrt{4\beta^{2}+1}-1\right)  ^{2}+4\beta
^{2}\right]  ^{-1}.$

Now, we evaluate the phases developed by the spinors $\Psi_{n,\lambda
}^{\left(  \mu\right)  }$: the dynamic phase (dyn), the Aharonov-Anandan (AA)
geometrical phase, and the total one. We first determine the phases associated
with the eigenspinor $\Psi_{n,\lambda}^{\left(  1\right)  }$. The AA
geometrical phase \cite{AA,Oh} is given by
\begin{equation}
\Phi_{AA}^{\left(  1\right)  }=i\int_{0}^{2\pi}\left[  \tilde{\Psi}^{\left(
1\right)  }\left(  \varphi\right)  \right]  ^{\dagger}\text{ }\frac
{d\tilde{\Psi}^{\left(  1\right)  }\left(  \varphi\right)  }{d\varphi}%
d\varphi,
\end{equation}
which leads to
\begin{equation}
\Phi_{AA}^{\left(  1\right)  }=2\pi\sin^{2}\left(  \theta/2\right)
.\label{H.24}%
\end{equation}
On the other hand, the dynamic phase \cite{Berry,Oh,Choi} is given by
\begin{equation}
\Phi_{dyn}^{\left(  1\right)  }=-\int_{0}^{2\pi}\left[  \tilde{\Psi}^{\left(
1\right)  }\left(  \varphi\right)  \right]  ^{\dagger}\text{ }H_{eff}%
\tilde{\Psi}^{\left(  1\right)  }\left(  \varphi\right)  d\varphi,
\end{equation}
involving
\begin{equation}
H_{eff}=-\left[  \beta_{11}\left(  \tilde{\sigma}_{\rho}\right)  -\beta
_{12}\left(  \tilde{\sigma}_{\varphi}\right)  \right]  ,
\end{equation}
that takes on the matrix form
\begin{equation}
H_{eff}=\left[
\begin{array}
[c]{cc}%
0 & \beta e^{i\left(  \delta+\varphi\right)  }\\
\beta e^{-i\left(  \delta+\varphi\right)  } & 0
\end{array}
\right]  .\label{H.26}%
\end{equation}
The evaluation implies%
\begin{equation}
\Phi_{dyn}^{\left(  1\right)  }=-2\pi\beta\sin\theta.\label{H.27}%
\end{equation}

The total phase,$\Phi_{Total}^{\left(  1\right)  }=\Phi_{AA}^{\left(
1\right)  }+\Phi_{dyn}^{\left(  1\right)  },$ acquired by the spinor,
$\Psi_{n,\lambda}^{\left(  1\right)  }$, is:%
\begin{equation}
\Phi_{Total}^{\left(  1\right)  }=-\pi\left(  1-\sqrt{1+4\beta^{2}}\right)
.\label{H.29}%
\end{equation}

Repeating all evaluations for the spinor $\Psi_{n,\lambda}^{\left(  2\right)
},$ in accordance with the definitions,
\begin{equation}
\Phi_{AA}^{\left(  2\right)  }=i\int_{0}^{2\pi}\left[  \tilde{\Psi}^{\left(
2\right)  }\left(  \varphi\right)  \right]  ^{\dagger}\text{ }\frac
{d\tilde{\Psi}^{\left(  2\right)  }\left(  \varphi\right)  }{d\varphi}%
d\varphi,
\end{equation}%
\begin{equation}
\Phi_{dyn}^{\left(  2\right)  }=-\int_{0}^{2\pi}\left[  \tilde{\Psi}^{\left(
2\right)  }\left(  \varphi\right)  \right]  ^{\dagger}\text{ }H_{eff}%
\tilde{\Psi}^{\left(  2\right)  }\left(  \varphi\right)  d\varphi,
\end{equation}
the generated phases are%
\begin{align}
\Phi_{AA}^{\left(  2\right)  } &  =2\pi\cos^{2}\left(  \theta/2\right)  ,\\
\Phi_{dyn}^{\left(  2\right)  } &  =2\pi\beta\sin\theta.
\end{align}

The total phase, $\Phi_{Total}^{\left(  2\right)  }=\Phi_{AA}^{\left(
2\right)  }+\Phi_{dyn}^{\left(  2\right)  },$ is
\begin{equation}
\Phi_{Total}^{\left(  2\right)  }=\pi\left(  1+\sqrt{4\beta^{2}+1}\right)
.\label{H.34}%
\end{equation}

Now, we note that it is possible to write the results (\ref{H.29}) and
(\ref{H.34}) as
\begin{equation}
\Phi_{Total}^{\left(  \mu\right)  }=\left(  -1\right)  ^{\mu}\pi\left[
1+\left(  -1\right)  ^{\mu}\sqrt{1+4\beta^{2}}\right]  ,\label{PhT1}%
\end{equation}
with $\beta$ given by Eq. (\ref{betaLV2}). Note that it coincides with Eq.
(\ref{Total1}), justifying the expression (\ref{H.14}) for the eigenenergies.
Hence, we have shown that the Lorentz-violating term $d^{ij}\sigma^{i}p^{j}$
generates, in the entire absence of electromagnetic field, geometrical and
total phases analogue to the ones provided by the Rashba coupling in condensed
matter systems, as previously discussed in Refs.
\cite{Oh,Choi,Molnar,Berche,Frus}.

In the case the time reversal $\left(  T\right)  $\ is a symmetry of the
system, there exists no persistent charge current and only persistent spin
current can arise \cite{Oh,Zhou,Berche}. More details about persistent
currents can be found in Refs. \cite{Splett,Sun}. As the parameters
$d^{00},d^{ij}\ $are $T-$even and the Hamiltonian (\ref{Hdij3}) is
$T-$symmetric, a nonnull spin current may be induced \cite{Berche}, being
defined as\textbf{\ }%
\begin{equation}
\mathcal{J}_{z}=\Psi_{n,\lambda}^{\left(  \mu\right)  \dag}\left\{
\mathbf{v}_{\varphi}\mathbf{,s}_{z}\right\}  \Psi_{n,\lambda}^{\left(
\mu\right)  },
\end{equation}
where $\Psi_{n,\lambda}^{\left(  \mu\right)  }$ is the spinor (\ref{Spinor1a})
and v$_{\varphi}=ir_{0}\left[  H,\varphi\right]  $\ is the azimuthal velocity
along the ring,\textbf{\ }%
\begin{equation}
\mathrm{v}_{\varphi}=-\frac{i}{mr_{0}}\partial_{\varphi}-\left(  d_{11}%
\tilde{\sigma}_{\rho}-d_{12}\tilde{\sigma}_{\varphi}\right)  .
\end{equation}
\textbf{\ }The measurable current is a kind of average on the degenerate
states of the system \cite{Splett} divided by the dimension of the system,
that is,
\begin{equation}
I_{z}=\frac{1}{2\pi r_{0}}\left\langle \mathcal{J}_{z}\right\rangle
,\label{MesC}%
\end{equation}
with
\begin{align}
\left\langle \mathcal{J}_{z}\right\rangle  &  =\left[  \Psi_{1,\text{ }%
+}^{\left(  2\right)  }\right]  ^{\dagger}\left\{  \mathbf{v}_{\varphi
}\mathbf{,s}_{z}\right\}  \Psi_{1,\text{ }+}^{\left(  2\right)  }+\left[
\Psi_{0,\text{ }-}^{\left(  1\right)  }\right]  ^{\dagger}\left\{
\mathbf{v}_{\varphi}\mathbf{,s}_{z}\right\}  \Psi_{0,\text{ }-}^{\left(
1\right)  }\nonumber\\
&  +\left[  \Psi_{-1,\text{ }-}^{\left(  2\right)  }\right]  ^{\dagger
}\left\{  \mathbf{v}_{\varphi}\mathbf{,s}_{z}\right\}  \Psi_{-1,\text{ }%
-}^{\left(  2\right)  }+\left[  \Psi_{0,\text{ }+}^{\left(  1\right)
}\right]  ^{\dagger}\left\{  \mathbf{v}_{\varphi}\mathbf{,s}_{z}\right\}
\Psi_{0,\text{ }+}^{\left(  1\right)  },\label{SpinC1}%
\end{align}
being the average spin current carried out on the four degenerate states,
$\Psi_{0,\text{ }-}^{\left(  1\right)  },\Psi_{0,\text{ }+}^{\left(  1\right)
},\Psi_{1,\text{ }+}^{\left(  2\right)  },\Psi_{-1,\text{ }-}^{\left(
2\right)  }$, with energy $\frac{1}{4}\left[  1-\sqrt{1+4\beta^{2}}\right]
^{2}.$

Performing the evaluation (\ref{SpinC1}) with
\begin{equation}
\left\{  \mathbf{v}_{\varphi}\mathbf{,s}_{z}\right\}  =\frac{1}{mr_{0}}\left[
\begin{array}
[c]{cc}%
-i\partial_{\varphi} & 0\\
0 & i\partial_{\varphi}%
\end{array}
\right]  ,
\end{equation}
we obtain the spin current density:%
\begin{equation}
\mathcal{J}_{z}=\frac{2}{mr_{0}}\left(  \cos\theta-1\right)  .
\end{equation}
Considering the expression (\ref{H.23b}), and expanding $\cos\theta$\ for
small $\beta^{2},$\ the measurable current is%
\begin{equation}
I_{z}=\frac{1}{\pi mr_{0}^{2}}\left(  \cos\theta-1\right)  \simeq\frac
{-2\beta^{2}}{\pi mr_{0}^{2}},
\end{equation}
which can be properly used to constrain the magnitude of the $d_{ij}%
$\ coefficients. \ For a consistence issue, it is possible to demonstrate that
the charge current, $\Psi_{n,\lambda}^{\left(  \mu\right)  \dag}$v$_{\varphi
}\Psi_{n,\lambda}^{\left(  \mu\right)  },$ evaluated over the degenerated
states regarded in Eq. (\ref{SpinC1}), is really null, as expected.

\subsection{Phases and spin currents generated by the coefficient $d_{00}$}

It is possible to show that similar effects are induced by the term
$d_{00}{\sigma}^{j}p^{j}$ of the Hamiltonian (\ref{Hdij2}), written as
\begin{equation}
d_{00}{\sigma}^{j}p^{j}=d_{00}\left(  \sigma_{x}p_{x}+\sigma_{y}p_{y}\right)
.
\end{equation}
In this case, we set $d_{ij}=0$, \ and $d_{33}=-d_{00},$\ in (\ref{Hdij2}), so
that we consider the simple Hamiltonian
\begin{equation}
H=\frac{\mathbf{p}^{2}}{2m}+d_{00}\sigma^{i}p^{i},
\end{equation}
which for an electron in a constant radius ring, $r=r_{0},$ in polar
coordinates, is \textbf{\ }%
\begin{equation}
H=\frac{1}{2mr_{0}^{2}}\left[  i\frac{\partial}{\partial\varphi}\right]
^{2}+\alpha_{00}\left(  \sigma_{x}\sin\varphi-\sigma_{y}\cos\varphi\right)
\left[  i\frac{\partial}{\partial\varphi}\right]  ,
\end{equation}
for $\alpha_{00}=\left(  1/r_{0}\right)  d^{00}.$ The hermitian form of this
Hamiltonian is
\begin{equation}
H=\Omega\left[  i\frac{\partial}{\partial\varphi}\right]  ^{2}+\alpha_{00}%
\bar{\sigma}_{\rho}\left[  i\frac{\partial}{\partial\varphi}\right]  +\frac
{i}{2}\alpha_{00}\bar{\sigma}_{\varphi}.\label{H002}%
\end{equation}
where
\begin{align}
\bar{\sigma}_{\rho} &  =\sigma_{x}\sin\varphi-\sigma_{y}\cos\varphi,\\
\bar{\sigma}_{\varphi} &  =\sigma_{x}\cos\varphi+\sigma_{y}\sin\varphi.
\end{align}
The Hamiltonian (\ref{H002}) can be read as a squared form
\begin{equation}
H=\Omega\left[  i\frac{\partial}{\partial\varphi}+\frac{1}{2\hslash\Omega
}\alpha_{00}\bar{\sigma}_{\rho}\right]  ^{2},\label{H00}%
\end{equation}
except by the term $\alpha_{00}^{2}/\left(  2\Omega\right)  ^{2}.$ Observing
that $H=\Omega H^{^{\prime\prime}2},$ the Hamiltonian (\ref{H00}) is expressed
as
\begin{equation}
H^{^{\prime\prime}}=%
\begin{bmatrix}
i\frac{\partial}{\partial\varphi} & i\beta_{00}e^{-i\varphi}\\
-i\beta_{00}e^{i\varphi} & i\frac{\partial}{\partial\varphi}%
\end{bmatrix}
,
\end{equation}
where
\begin{equation}
\beta_{00}=\left(  mr_{0}\right)  d_{00}.\label{Beta00}%
\end{equation}
The eigenenergies are%
\begin{equation}
E_{n}=\Omega\left\{  \lambda n+\frac{1}{2}\left[  1-\left(  -1\right)  ^{\mu
}\sqrt{1+4\beta_{00}^{2}}\right]  \right\}  ^{2},
\end{equation}%
\begin{equation}
E_{n}=\Omega\left[  \lambda n-\frac{\Phi_{Total}^{\left(  \mu\right)  }}{2\pi
}\right]  ^{2},
\end{equation}
with $n\in\mathbb{Z}$, $\lambda=\pm1,$ and
\begin{equation}
\Phi_{Total}^{\left(  \mu\right)  }=-\pi\left[  1-\left(  -1\right)  ^{\mu
}\sqrt{1+4\beta_{00}^{2}}\right]  .\label{PhT2}%
\end{equation}
The eigenspinors have the general form
\begin{equation}
\chi_{n,\lambda}^{\left(  \mu\right)  }=e^{i\lambda n\varphi}\tilde{\chi
}^{\left(  \mu\right)  }\left(  \varphi\right)  ,\label{H.21}%
\end{equation}
and explicit solution
\begin{align}
\tilde{\chi}^{\left(  1\right)  }\left(  \varphi\right)   &  =\left[
\begin{array}
[c]{c}%
\cos\left(  \vartheta/2\right) \\
i\sin\left(  \vartheta/2\right)  e^{i\varphi}%
\end{array}
\right]  ,\\
\tilde{\chi}^{\left(  2\right)  }\left(  \varphi\right)   &  =\left[
\begin{array}
[c]{c}%
\sin\left(  \vartheta/2\right) \\
-i\cos\left(  \vartheta/2\right)  e^{i\varphi}%
\end{array}
\right]  ,
\end{align}
with $\mu=1$ or $\mu=2$ representing spin down and spin up, respectively, and
\begin{equation}
\tan\left(  \vartheta/2\right)  =\frac{1}{\beta_{00}}\left[  \frac{1}{2}%
+\frac{1}{2}\sqrt{1+4\beta_{00}^{2}}\right]  ,
\end{equation}%
\begin{equation}
\sin^{2}\left(  \vartheta/2\right)  =\frac{\left(  1+\sqrt{4\beta_{00}^{2}%
+1}\right)  ^{2}}{\left(  \sqrt{4\beta_{00}^{2}+1}+1\right)  ^{2}+4\beta
_{00}^{2}},
\end{equation}%
\begin{equation}
\cos\vartheta=-1/\sqrt{1+4\beta_{00}^{2}},
\end{equation}

Following the same steps of the first case, we obtain the AA and dynamic
phases for the spinor $\tilde{\Psi}^{\left(  1\right)  }$ :%
\begin{align}
\Phi_{AA}^{\left(  1\right)  } &  =-2\pi\sin^{2}\left(  \vartheta/2\right)
,\\
\Phi_{dyn}^{\left(  1\right)  } &  =-2\pi\beta_{00}\sin\vartheta,
\end{align}
and for the spinor $\tilde{\Psi}^{\left(  2\right)  }:$%
\begin{align}
\Phi_{AA}^{\left(  2\right)  } &  =-2\pi\cos^{2}\left(  \vartheta/2\right)
,\label{H.32}\\
\Phi_{dyn}^{\left(  2\right)  } &  =2\pi\beta_{00}\sin\vartheta,
\end{align}
implying the total phases%
\begin{align}
\Phi_{Total}^{\left(  1\right)  } &  =-\pi\left(  1+\sqrt{1+4\beta_{00}^{2}%
}\right)  ,\\
\Phi_{Total}^{\left(  2\right)  } &  =-\pi\left(  1-\sqrt{1+4\beta_{00}^{2}%
}\right)  ,
\end{align}
which are in the pattern of expression (\ref{PhT2}). Thus, we conclude that
the coefficient $d_{00}$ also succeeds in generating geometrical and total
phases to electrons in 1-dim rings similar to the ones provided by the Rashba
or Dresselhauss interactions.

The measurable current stems from Eq. (\ref{MesC}), with spinors
$\chi_{n,\lambda}^{\left(  \mu\right)  }$ given Eq. (\ref{H.21}) and the
average density current explicitly written as
\begin{align}
\left\langle \mathcal{J}_{z}\right\rangle  &  =\left[  \chi_{0,\text{ }%
+}^{\left(  2\right)  }\right]  ^{\dagger}\left\{  \mathbf{v}_{\varphi
}\mathbf{,s}_{z}\right\}  \chi_{0,\text{ }+}^{\left(  2\right)  }+\left[
\chi_{0,\text{ }-}^{\left(  2\right)  }\right]  ^{\dagger}\left\{
\mathbf{v}_{\varphi}\mathbf{,s}_{z}\right\}  \chi_{0,\text{ }-}^{\left(
2\right)  }\nonumber\\
&  +\left[  \chi_{1,\text{ }-}^{\left(  1\right)  }\right]  ^{\dagger}\left\{
\mathbf{v}_{\varphi}\mathbf{,s}_{z}\right\}  \chi_{1,\text{ }-}^{\left(
1\right)  }+\left[  \chi_{-1,\text{ }+}^{\left(  1\right)  }\right]
^{\dagger}\left\{  \mathbf{v}_{\varphi}\mathbf{,s}_{z}\right\}  \chi
_{-1,\text{ }+}^{\left(  1\right)  },
\end{align}
while $\chi_{0,\text{ }-}^{\left(  2\right)  },\chi_{0,\text{ }+}^{\left(
2\right)  },\chi_{1,\text{ }-}^{\left(  1\right)  },\chi_{-1,\text{ }%
+}^{\left(  1\right)  }$ are the four degenerate states with energy $\frac
{1}{4}\left[  1-\sqrt{1+4\beta_{00}^{2}}\right]  ^{2}.$

The evaluation (\ref{SpinC1}) leads to the following spin current density:%
\begin{equation}
\mathcal{J}_{z}=\frac{2}{mr_{0}}\left(  1+\cos\vartheta\right)  ,
\end{equation}
related to the corresponding persistent spin current
\begin{equation}
I_{z}=\frac{1}{\pi mr_{0}^{2}}\left(  1+\cos\vartheta\right)  \simeq
-\frac{2\beta_{00}^{2}}{\pi mr_{0}^{2}},
\end{equation}
or $I_{z}=-2md_{00}^{2}/\pi,$ to be used in the next section to impose upper
limits on the magnitude of the LV\ parameters.

\section{Upper bounds on the LV parameters}

An interesting issue concerns the use of these results to obtain upper bounds
on the magnitude of the Lorentz-violating coefficients, $d^{ij},d^{00},$
responsible for the effects here reported. In accordance with Ref.
\cite{KostelecR}, the best upper bounds on the coefficients $d^{ij} $ reach
the level of 1 part in $10^{14}-10^{15},$ being achieved by means of TeV
inverse Compton radiation from astronomical sources \cite{AltschulR}. There
are not tight bounds at all for these coefficients in the context of condensed
matter systems. In Ref. \cite{Ajaib2}, there were estimated upper bounds of
the order $d^{ij}<$ $10^{-2}$ by straightforward comparison with the Rashba
constant magnitude in electronic systems. However, the phenomenology of
electrons in 1-dim rings can provide some mechanisms that can "amplify" the
Lorentz-violating effects, leading to much better upper bounds. There are at
least two main routes to do it: one involving measurements of topological
phases, other related to measurements of persistent spin currents.

The first route to set upper limits in the LV parameters is analyzing the
geometrical phases induced in the absence of fields. Since 1989 it is
experimentally known the possibility of measuring A-Casher phases as small as
$10^{-3}$ rad \cite{ACM}, \cite{Chandra}. There are also several works
analyzing the conductance {oscillations and transport properties in 1-dim
rings endowed with induced A-Casher phase
\cite{Choi,Molnar,Nitta,Shen,Frus,Kovalev,Konig}. As in the present work the
LV terms predict phase and persistent current induction independently of
electromagnetic fields, a suitable experiment to constrain such terms should
investigate the generation of geometrical phase and correlated effects on
1-dimensional electrons in the absence of spin-orbit couplings and
electromagnetic fields.}

We should first consider the total phase yielded by the coefficient $d^{00},$
given by Eq. (\ref{PhT2}). Taking $\mu=1$ and considering $\beta_{00}^{2}<<1,$
we have $\left\vert \Phi_{Total}^{\left(  \mu\right)  }\right\vert \simeq
2\pi\beta_{00}^{2}=2\pi\left(  mr_{0}\right)  ^{2}d_{00}^{2},$ in accordance
with Eq. (\ref{Beta00}). For electrons of effective mass $m=0.05m_{e}$ in a
typical mesoscopic ring with $r_{0}=1\mu m, $ it holds $mr_{0}=1.3\times
10^{5}$ \cite{Konig}, \cite{Chandra}. If we consider a mesoscopic ring in the
absence of fields and spin-orbit couplings, no phases can be generated at all,
so that we state $\left\vert \Phi_{Total}^{\left(  \mu\right)  }\right\vert
<10^{-3}$ rad. This general condition leads to:
\begin{equation}
\left\vert d_{00}\right\vert <1.0\times10^{-7}.
\end{equation}
standing for a bound that can be communicated for the components $d^{ij},$
once the tensor $d^{\mu\nu}$ is traceless.\ In the configuration of
interest,\textbf{\ }$\left\vert d_{00}\right\vert =\left\vert d_{33}%
\right\vert $\textbf{. }This bound is not so good as the ones of Refs.
\cite{KostelecR,AltschulR}, but we should remark that it is now estimated in
the context of a mesoscopic condensed matter system, being much better than
the best one suggested in Ref. \cite{Ajaib2}, $d^{ij}\sim$ $10^{-2}$, by a
factor $10^{5}.$ The same procedure can be applied to the total phase
(\ref{PhT1}) provided by the coefficient $d_{ij},$ implying the same result
$\left\vert d_{ij}\right\vert <10^{-7}.$ As the upper bound is proportional to
$1/r_{0},$ increasing the radius leads to tighter limits, obviously without
loosing the mesoscopic character of the system.

Another effective way to constrain the LV parameters is appealing to the spin
persistent current associated with this model. It is known that currents as
small as $0.1$ nA can be measured in mesoscopic rings \cite{Chandra}. Noting
that persistent charge and spin currents can be distinguished from each other
\cite{Splett,Sun}, working in a 1-dimension ring endowed with $T-$symmetry in
the absence of fields and spin-orbit couplings, no spin persistent current can
be generated at all. This \textit{gedanken }system allows to impose
$2md_{00}^{2}/\pi<10^{-10}$A, which yields
\[
d_{00}<1.4\times10^{-6},
\]
and \ similarly, $\left\vert d_{33}\right\vert <1.4\times10^{-6}.$ \ Here, we
used $1A=3.52\times10^{2}eV.$ This second route yields a level of constraining
one order of magnitude below that latter one but seems to be more confident
and realizable, once current measurements are much more precise and accessible
than the ones of phases. Even in this case, the bound remains at least four
order of magnitude better than the ones estimated in Ref. \cite{Ajaib2} by
direct comparison with the Rash/Dresselhauss couplings.

\section{Conclusions}

In this work, we have shown that LV terms belonging to the fermion sector of
the SME can induce geometrical phases to electrons in 1-dim rings similar to
the ones yielded by the Rashba and Dresselhauss interactions in condensed
matter systems, in total absence of electromagnetic fields. Particularly, we
analyzed the nonrelativistic terms stemming from $\overline{\Psi}d^{\mu\nu
}\gamma_{5}\gamma_{\mu}\Psi$, belonging to the fermion sector. \ After
carrying out the modified eigenenergies, we have used the evaluated
eigenspinors for explicitly computing the geometrical and dynamic phases
developed by the electrons. The phases achieved present a similar pattern to
the ones induced by the Rashba coupling. The key role of the (absent) electric
field is now played by the LV background. We also carried out the spin
persistent currents induced by these coefficients, using it to impose the
upper bound $d_{00}<1.4\times10^{-6},$ the best one obtained in the context of
mesoscopic system until the moment.

Lorentz-violating effects here reported can be also induced by other terms of
the SME fermion sector. Analyzing the full nonrelativistic limit of the SME
fermion sector, see Refs. \cite{Fermion2}, we focus on the following
Hamiltonian terms,%
\begin{equation}
\frac{1}{2}\epsilon_{klm}g_{mlj}p^{j}\sigma^{k},\text{ }\frac{1}{2}%
\epsilon_{kjm}g_{m00}p^{j}\sigma^{k},\text{ }\epsilon_{jkl}\frac{H_{l0}}%
{m}p^{j}\sigma^{k},
\end{equation}
that possess the Rashba or Dresselhauss-like form. The first two ones yield
analogue effects to the ones ascribed to the coefficient $d_{ij}.$ We should
still comment that other terms of the nonrelativistic Hamiltonian could induce
similar geometrical phases, but with smaller magnitude by the factor $m^{-1}$.
The Rashba-like contribution of the nonminimal model \cite{NM3} can also find
similar role as the aforementioned terms, but in the presence of
electromagnetic field.

\begin{acknowledgments}
The authors thank to CNPq, CAPES, FAPEMA (Brazilian research agencies) and
FAPEMA/APCINTER-00256/14 for financial support. M.M Ferreira Jr is grateful to
FAPEMA/Universal/00836/13, CNPq/PQ 308852/2011-7,
CNPq/Universal/460812/2014-9. R. Casana is grateful to
CNPq/Universal/483863/2013-0, FAPEMA/Universal/Universal 00760/13, CNPq/PQ
311036/2012-0. E.O. Silva is grateful to CNPq/Universal/ 482015/2013-6,
CNPq/PQ 306068/2013-3 and FAPEMA/Universal/00845/13.
\end{acknowledgments}


\begin{thebibliography}{99}                                                                                               %
\bibitem {Colladay}D. Colladay, V. A. Kostelecky, Phys. Rev. D 55 (1997) 6760;
Phys. Rev. D 58 (1998) 116002 ; S. R. Coleman and S. L. Glashow, Phys. Rev. D
59 (1999) 116008; J. D. Tasson, Rep. Prog. Phys. 77 (2014) 062901.

\bibitem {Jackiw}S.M. Carroll, G.B. Field and R. Jackiw, Phys. Rev.\textit{\ }%
D 41 (1990) 1231.

\bibitem {Mewes}V. A. Kostelecky and M. Mewes, Phys. Rev. Lett. 87 (2001)
251304; Phys. Rev. Lett. 97 (2006) 140401; Phys. Rev. D 66, (2002) 056005; M.
Schreck, Phys. Rev. D 89 (2014) 085013; C. Hernaski, Phys. Rev. D 90 (2014) 124036.

\bibitem {Soldati}A.A. Andrianov and R. Soldati, Phys. Rev. D 51 (1995) 5961;
Phys. Lett. B 435 (1998) 449; J. Alfaro, A.A. Andrianov, M. Cambiaso, P.
Giacconi, R. Soldati, Int. J. Mod. Phys. A 25 (2010) 3271; J. Alfaro, A.A.
Andrianov, M. Cambiaso, P. Giacconi, R. Soldati, Phys. Lett. B 639 (2006) 586;
L. H. C. Borges, F. A. Barone, J. A. Helay\"{e}l-Neto, Eur. Phys. J. C 74
(2014) 2937; W. F. Chen and G. Kunstatter, Phys. Rev. D 62 (2000) 105029.

\bibitem {Adam}C. Adam and F. R. Klinkhamer, Nucl. Phys.\ B 607 (2001) 247;
Nucl. Phys.\ B 657 (2003) 214.

\bibitem {Baeta}A. P. Baeta Scarpelli, H. Belich, J. L. Boldo, J. A.
Helayel-Neto, Phys. Rev. D 67 (2003) 085021; H. Belich, J. L. Boldo, L. P.
Colatto, J.A. Helayel-Neto, A.L.M.A. Nogueira, Phys.Rev. D68 (2003) 065030; R.
Bufalo, Int. J. Mod. Phys. A 29 (2014) 1450112;

\bibitem {Vertex}F.R. Klinkhamer, M. Schreck, Nucl. Phys. B 848 (2011) 90; M.
Schreck, Phys. Rev. D\textbf{\ }86 (2012) 065038; M. A. Hohensee,\emph{\ et
al.}, Phys. Rev. D 80 (2009) 036010; A. Moyotl, H. Novales-S\'{a}nchez, J. J.
Toscano, E. S. Tututi, Int. J. Mod. Phys. A 29 (2014) 1450039; Int. J.
Mod.Phys. A 29 (2014) 1450107; M. Cambiaso, R. Lehnert, R. Potting, Phys. Rev.
D 90 (2014) 065003.

\bibitem {HD}V. A. Kostelecky and M. Mewes, Phys. Rev. D 80 (2009) 015020; M.
Cambiaso, R. Lehnert, R. Potting, Phys. Rev. D 85 (2012) 085023; M. Schreck,
Phys. Rev. D 89 (2014) 105019; Phys. Rev. D 90 (2014) 085025; B. Agostini, F.
A. Barone, F. E. Barone, P. Gaete, J. A. Helay\"{e}l-Neto, Phys. Lett. B 708
(2012) 212; L. Campanelli, Phys. Rev. D 90 (2014) 105014; R. Bufalo, B.M.
Pimentel, D.E. Soto, Physical Review D 90 (2014) 085012.

\bibitem {Reyes}C. M. Reyes, L. F. Urrutia, J. D. Vergara, Phys. Rev. D 78
(2008) 125011; Phys. Lett. B 675 (2009) 336; C. M. Reyes, Phys. Rev. D 82
(2010) 125036; Phys. Rev. D 80 (2009) 105008; Phys. Rev. D 87 (2013) 125028.

\bibitem {Fermion1}R. Bluhm, V.A. Kostelecky, N. Russel, Phys. Rev. Lett. 79
(1997) 1432; Phys. Rev. Lett. 82 (1999) 2254; Phys. Rev. D 57 (1998) 3932; R.
Bluhm, V.A. Kostelecky, C.D. Lane, Phys. Rev. Lett. 84 (2000) 1098.

\bibitem {Fermion2}V.A. Kostelecky and C. D. Lane, J. Math. Phys. (N.Y.) 40
(1999) 6245; V.A. Kostelecky, C.D. Lane, Phys. Rev. D 60\textbf{,} (1998) 116010.

\bibitem {Fermion3}V.A. Kostelecky and R. Lehnert, Phys. Rev. D 63 (2001)
065008; R. Lehnert, J. Math. Phys. (N.Y.) 45 (2004) 3399; R. Lehnert, Phys.
Rev. D 68 (2003) 085003; B. Goncalves, Y. N. Obukhov, Ilya L. Shapiro, Phys.
Rev. D 80 (2009) 125034; B. Gon\c{c}alves, M. M. Dias Junior, and B. J.
Ribeiro, Phys. Rev. D 90 (2014) 085026.

\bibitem {DiracE}R. V. Maluf, J. E. G. Silva, W. T. Cruz, C. A. S. Almeida,
Phys. Lett. B 738 (2014) 341; S. I. Kruglov, Phys. Lett. B. 718 (2012) 228;
Int. J. Mod. Phys. A 29 (2014) 1450031.

\bibitem {KostelecR}V.A. Kostelecky and N. Russel, Rev. Mod. Phys. 83, 11 (2011).

\bibitem {AltschulR}B. Altschul, Phys. Rev. D 75 (2007) 041301 (R).

\bibitem {Berry}M. V. Berry, Proceedings of the Royal Society of London 392
(1984) 45.

\bibitem {AA}Y. Aharonov and J. Anandan, Phys. Rev. Lett. 58 (1987) 1593.

\bibitem {Aronov1}A. G. Aronov, Y. V. Sharvin, Rev. Mod. Phys. 59 (1987) 755.

\bibitem {Loss}D. Loss, P. M. Goldbart, A. V. Balatsky, Phys. Rev. Lett. 65
(1990) 1655; D. Loss, P. M. Goldbart, Phys. Lett. A 215 (1996) 197.

\bibitem {Mathur}H. Mathur, A. D. Stone, Phys. Rev. Lett. 68 (1992) 2964 .

\bibitem {AC}Y. Aharonov and A. Casher, Phys. Rev. Lett. 53 (1984) 319 .

\bibitem {Balatsky}A. V. Balatsky, B. L. Altshuler, Phys. Rev. Lett. 70 (1993) 1678.

\bibitem {Aronov2}A. G. Aronov and Y. B. Lyanda-Geller, Phys. Rev. Lett. 70
(1993) 343.

\bibitem {Qian}T.-Z. Qian, Z.-B. Su, Phys. Rev. Lett. 72 (1994) 2311.

\bibitem {Oh}S. Oh, C.-M. Ryu, Phys. Rev. B 51 (1995) 13441.

\bibitem {Choi}T. Choi, S.Y. Cho, C-M. Ryu, and C. K. Kim, Phys. Rev. B 56
(1997) 4825.

\bibitem {Molnar}B. Molnar, F. M. Peeters, P. Vasilopoulos, Phys. Rev. B 69
(2004) 155335.

\bibitem {Meijer}F. E. Meijer, A. F. Morpurgo, T. M. Klapwijk, Phys. Rev. B 66
(2002) 033107.

\bibitem {Zhou}Y.-C. Zhou, H.-Z. Li, \ X. Xue, Phys. Rev. B 49 (1994) 14010;
S-L Zhu, Y.-C. Zhou, and H.-Z. Li, Phys. Rev. B 52 (1995) 7814.

\bibitem {Nitta}J. Nitta, T. Bergsten, \ New J. Phys. 9 (2007) 341 .

\bibitem {Berche}B. Berche, C. Chatelain, E. Medina, Eur. J. Phys. 31 (2010) 1267.

\bibitem {Gigl}S. Giglberger \textit{et al.}, Phys. Rev. B 75 (2007) 035327.

\bibitem {Splett}J. Splettst\"{o}sser, M. Governale, U. Z\"{u}licke, Phys.
Rev. B 68 (2003) 165341.

\bibitem {Shen}S.-Q. Shen, Z.-J. Li, Z. Ma, Appl. Phys. Lett. 84 (2004) 6.

\bibitem {Frus}D. Frustaglia, K. Richter, Phys. Rev. B 69 (2004) 235310.

\bibitem {Citro}R. Citro, F. Romeo, M. Marinaro, Phys. Rev. B 74 (2006) 115329.

\bibitem {Kovalev}A. A. Kovalev, Phys. Rev. B 76 (2007) 125307.

\bibitem {Nonminimal}H. Belich, T. Costa-Soares, M. M. Ferreira, Jr., and J.
A. Helayel-Neto, Eur. Phys. J. C 41 (2005) 421 .

\bibitem {Nonminimal2}H. Belich, E. O. Silva, M. M. Ferreira, Jr., and M. T.
D. Orlando, Phys. Rev. D 83 (2011) 125025.

\bibitem {Knut}K. Bakke, E. O. Silva, H. Belich, J. Phys. G, Nuclear and
Particle Physics 39 (2012) 055004; J. Math. Phys. 52 (2011) 063505; Annalen
der Physik (Leipzig) 523 (2011) 910; K. Bakke and H. Belich, J. Phys. G,
Nuclear and Particle Physics 39 (2012) 085001; Phys. Rev. D 90 (2014) 025026;
E. O. Silva, F. M. Andrade, H. Belich, C. Filgueiras, Eur. Phys. J. C 73
(2013) 2402.

\bibitem {Ajaib2}M. A. Ajaib, Lorentz violation and Condensed Matter Physics, arXiv1403.7622.

\bibitem {NM3}R. Casana, M. M. Ferreira, Jr., E. Passos, F. E. P. dos Santos,
and E. O. Silva, Phys. Rev. D 87 (2013) 047701 .

\bibitem {Bakke}K. Bakke and H. Belich, Ann. Phys. (Berlin) 526 (2013) 187; K.
Bakke and H. Belich, Ann. Phys. (Print) 354 (2015) 1.

\bibitem {Ajaib}M. A. Ajaib, International Journal of Modern Physics A 27
(2012) 1250139 .

\bibitem {Sun}Q.-f. Sun, X.C. Xie, J. Wang, Phys. Rev. B 77 (2008) 035327.

\bibitem {ACM}A. Cimmino, G. I. Opat, A. G. Klein, H. Kaiser, S.A. Werner, M.
Arif, and R. Clothier, Phys. Rev. Lett. 63 (1989) 380; K. Sangster, E.A.
Hinds, S. M. Barnett, and E. Riis, Phys. Rev. Lett. 71, 3641 (1993); A.
G\"{o}rlitz, B.Schuh, and A. Weis, Phys. Rev. A 51 (1995) R4305.

\bibitem {Konig}M. K\"{o}nig, A. Tschetschetkin, E. M. Hankiewicz, J. Sinova,
V. Hock, V. Daumer, M. Sch\"{a}fer, C. R. Becker, H. Buhmann, and L.W.
Molenkamp, Phys. Rev. Lett. 96 (2006) 076804; T. Bergsten, T. Kobayashi, Y.
Sekine, and J. Nitta, Phys. Rev. Lett. 97 (2006) 196803.

\bibitem {Chandra}V. Chandrasekhar, R. A. Webb, M. J. Brady, M. B. Ketchen, W.
J. Gallagher, and A. Kleinsasser, Phys. Rev. Lett. 67 (1991) 3578; D. Mailly,
C. Chapelier, and A. Benoit, Phys. Rev. Lett. 70 (1993) 2020.
\end{thebibliography}
\end{document}